\documentclass[a4paper,journal]{IEEEtran}
\usepackage{mathrsfs}
\usepackage[dvips]{graphicx}
\usepackage{subfigure}
\usepackage{color}
\usepackage{url}
\usepackage[noadjust]{cite}
\usepackage{multicol}
\usepackage{amssymb,dsfont,amsmath,amsfonts,mathrsfs,soul}
\usepackage{xspace,colortbl,graphicx}
\usepackage{enumerate}
\usepackage[ruled]{algorithm2e}
\usepackage{color, soul} 
\usepackage{multirow}
\usepackage{arydshln}
\usepackage{pifont}
\usepackage{mathtools} 
\usepackage{cases}
\usepackage{framed}

\newtheorem{theorem}{\indent Theorem}[section]

\newtheorem{definition}{\indent Definition}[section]
\newtheorem{remark}{\indent Remark}[section]

\newcommand{\be}{\begin{equation}}
\newcommand{\ee}{\end{equation}}
\newcommand{\bea}{\begin{eqnarray}}
\newcommand{\eea}{\end{eqnarray}}
\newcommand{\ba}{\begin{array}}
\newcommand{\ea}{\end{array}}

\sethlcolor{yellow}

\ifCLASSINFOpdf

\else

\fi

\large
\begin{document}
\title{
Selection of Time Headway in Connected and Autonomous Vehicle Platoons under  Noisy V2V Communication 
}

\author{Guoqi~Ma, Prabhakar~R.~Pagilla$^\ast$, Swaroop~Darbha

\thanks{The authors are with the Department of Mechanical Engineering, Texas A\&M University, College Station, TX 77843, USA (e-mail: gqma@tamu.edu; ppagilla@tamu.edu; dswaroop@tamu.edu). $^\ast$Corresponding author.

Part of this work 
was presented at the 26th IEEE International Conference on Intelligent Transportation Systems (ITSC), Bilbao, Spain, 2023~\cite{ITSC2023}.

}}



\maketitle

	
\begin{abstract}
 In this paper, we investigate the selection of time headway to ensure robust string stability in connected and autonomous vehicle platoons in the presence of signal noise in Vehicle-to-Vehicle (V2V) communication. In particular, we consider the effect of noise in communicated vehicle acceleration from the predecessor vehicle to the follower vehicle on the selection of the time headway in predecessor-follower type vehicle platooning with a Constant Time Headway Policy (CTHP). Employing a CTHP based control law for each vehicle that utilizes on-board sensors for measurement of position and velocity of the predecessor vehicle and wireless communication network for obtaining the acceleration of the predecessor vehicle, we investigate how time headway is affected by communicated signal noise. We derive constraints on the CTHP controller gains for predecessor acceleration, velocity error and spacing error and a lower bound on the time headway which will ensure robust string stability of the platoon against signal noise. We provide comparative numerical simulations on an example to illustrate the main result. 
\end{abstract}


\section{Introduction} 
\IEEEPARstart{T}{he} deployment of connected and autonomous vehicle platoons has the potential to benefit transportation systems in a profound and comprehensive way~\cite{9989507,10026667}. Initially, under the Adaptive Cruise Control (ACC) paradigm facilitated by the use of onboard sensors (such as radars) for measuring the velocities of and distances from adjacent vehicles, the vehicle platoon can maintain a constant inter-vehicular spacing, referred to as Constant Spacing Policy (CSP)~\cite{10.1115/1.2802497}. Recently, the benefits of employing advanced vehicular communication technologies and modern communication protocols have been expounded in great detail in the literature; for example, Dedicated Short Range Communication (DSRC)~\cite{5888501}, Long Term Evolution (LTE)~\cite{8108463}, 5G~\cite{9345798}, and V2V communication~\cite{vinel2015vehicle}. In addition to the use of onboard sensors, advanced vehicular wireless communications can lead to a higher-level of connectivity by incorporating the acceleration information of other vehicles which has the potential to significantly improve safety, increase mobility and throughput, and reduce fuel consumption under the paradigm of Cooperative Adaptive Cruise Control (CACC) with Constant Time Headway Policy (CTHP)~\cite{darbha2018benefits}. 
  
In addition to internal stability for each vehicle, the connected and autonomous vehicle platoon should also exhibit robust string stability, i.e., robustness of the platoon to uncertain lag, noise, and external disturbances. In recent decades, there has been extensive research devoted to connected and autonomous vehicle platoons on a wide range of topics including controller design~\cite{6683051,SILVA2021109542,1105955,JIANG2021103110,SCHOLTE2022103511,doi:10.1287/trsc.2021.1100}, communication mechanisms~\cite{iet2020platooning,BIAN201987,9497781}, experimental validation in realistic environments~\cite{5571043,GE2018445}, mixed human and autonomous traffic~\cite{9246221,GONG201825,9626600}, 
string stability analysis~\cite{6515636,7879221}. 
However, most existing results assume ideal V2V communication which is often not practical. Among many other factors, signal noise in communication channels is a key concern, which can be caused by a variety of factors including limited communication  bandwidth~\cite{8606268}, cyber-attacks~\cite{9477412}, etc.
If key control parameters, such as time headway, are chosen without consideration of signal noise in communicated signals, then there is a possibility of the onset of string instability and collisions. Although communication quality has improved significantly under 5G and more advanced wireless communication, signal noise when propagating through the platoon could have a substantial affect on safety and performance. 

Previous research on imperfect communication focused on how packet drops affected string stability~\cite{9457141}. In~\cite{10018490,9914654}, the disturbance estimation and control problem was investigated, though without considering the quantitative effect of disturbance on the design parameters. In contrast to prior research with imperfect communication, this paper considers the following problem: Given a Signal-to-Noise Ratio (SNR) with V2V communication, what is the lower bound of the employable time headway for CACC based vehicle platooning? For this purpose, we consider a CTHP based control law where the feedforward acceleration of the predecessor vehicle is communicated and contains signal noise. Since the communication is through an $n$-bit channel, we model the noise in the acceleration signal as a sum of $n$ binary random variables. To address the issue of the vehicle closed-loop governing equation being stochastic due to the noise signal, we develop an equivalent deterministic governing equation using the averaging procedure described in~\cite{9457141}. Based on this, we obtain the spacing error propagation equation and show that the platoon is robustly string stable in the presence of signal noise and parasitic lag and derive conditions on control gains and time headway under which the platoon is robustly string stable. We provide numerical simulation results on a platoon to verify the main result. 

The main contributions of the work can be summarized as follows. In the context of communication from a single predecessor vehicle, this paper provides an extension to our previous work in~\cite{darbha2018benefits} by considering signal noise in communicated acceleration from the predecessor vehicle. We derive a lower bound on the time headway that is a function of the parasitic lag in the vehicle dynamics ($\tau_0$), acceleration gain ($k_a$), and SNR ($\rho$). This lower bound aids in the selection of the time headway in the presence of noise. Further, the lower bound reduces to the ideal communication case given in~\cite{darbha2018benefits} with no noise in the communicated acceleration signal. In addition, our formulation and analysis allow for systematic selection of the control gains as well as time headway for a given signal to noise ratio. 

The remainder of the paper is organized as follows. Section~\ref{section:problem-formulation-and-preliminaries} contains preliminaries including vehicle dynamics and relevant definitions. The main theoretical results are provided in Section~\ref{section:main-results}. An illustrative numerical example and simulation results are provided in Section~\ref{section:numerical-simulations}. Finally, some concluding remarks are given in Section~\ref{section:conclusion}.

\section{Preliminaries} \label{section:problem-formulation-and-preliminaries}
\begin{figure}[!htb]
\centering{\includegraphics[scale=0.60]{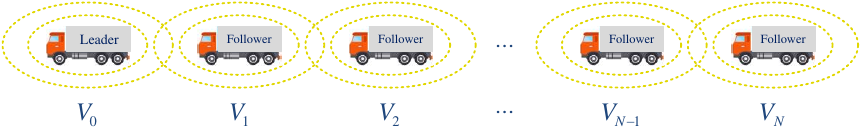}}
\caption{Autonomous and connected vehicle platoon with V2V communication.
\label{fig_1}}
\end{figure}

Consider a string of autonomous vehicles equipped with V2V communication as illustrated in Fig.~\ref{fig_1}.
The $i$-th vehicle dynamics is given by the model: 
\begin{align} \label{eq:vehicle-dynamics}
\left\{\begin{array}{*{20}{l}}
\ddot{x}_{i}(t) = a_{i}(t), \\
\tau \dot{a}_{i}(t) + a_{i}(t) = u_{i}(t),
\end{array}
\right.
\end{align}
where $x_{i}(t)$, $a_{i}(t)$, $u_{i}(t)$ represent the position, acceleration, and control input of the $i$-th vehicle at time instant $t$, and $i \in \mathcal{N} = \{ 1, 2, \cdots, N \}$, where $N$ is the total number of the following vehicles in the platoon, $\tau$ denotes the parasitic actuation lag. It is assumed that $\tau$ is {\it uncertain} with $\tau \in \left( 0, \tau_{0} \right]$, where $\tau_{0}$ is a positive real constant.

\begin{definition}
    Let $d$ denote the minimum or standstill spacing between adjacent vehicles, $v_i(t)$ denote the velocity of the $i$-th vehicle, $h_w$ denote the time headway, and $e_i(t) := x_{i}(t) - x_{i - 1}(t) + d$ be the spacing error for the $i$-th vehicle with respect to the $(i-1)$-th vehicle. Define the velocity dependent inter-vehicular spacing error for the $i$-th vehicle as: 
    \begin{align} \label{eq:delta-i-definition}
     \delta_i(t) = e_i(t) + h_w v_i(t).   
    \end{align}
\end{definition}
\begin{definition} [\cite{darbha2018benefits}] \label{definition:string-stability}
Let $\delta_i(s)$ denote the Laplace transform of $\delta_i(t)$ and let $H(s)$ denote the spacing error propagation transfer function that satisfies $\delta_{i}(s) = H(s) \delta_{i - 1}(s)$. 
The connected and autonomous vehicle platoon is said to be robustly string stable if the following two conditions hold for all $\tau\in (0,\tau_0]$: (i) $H(s)$ achieves internal stability and (ii) the platoon is string stable, i.e., it holds that $\Vert \delta_{i}(t) \Vert_{\infty} \le \Vert \delta_{i - 1}(t) \Vert_{\infty}$, or in the frequency domain
\begin{align} \label{eq:definition-2.1-H-j-omega-infinity-norm}
    \Vert H(j\omega) \Vert_{\infty} \le 1.
\end{align}
\end{definition}


\begin{definition}
Let $s(t)$ and $n(t)$ denote the transmitted signal and the noise signal, respectively, and assume 
$n(t) = 0$ when $s(t) = 0$. Then, the Signal-to-Noise Ratio (SNR) for $s(t) \neq 0$ is defined as 
 \begin{align} \label{eq:SNR-definition}
  {\rm SNR} = \min\limits_{t > 0} 20 \log \frac{\vert s(t) \vert}{\vert n(t) \vert},     
 \end{align}
where 
$\rm log$ denotes the logarithmic function with base 10.  From the ${\rm SNR}$ definition~\eqref{eq:SNR-definition}, we can obtain that if ${\rm SNR} \coloneqq \varrho$, then $\rho = \min\limits_{t > 0} \frac{\vert s(t) \vert}{\vert n(t) \vert} = 10^{\frac{\varrho}{20}}$, and for notational convenience, we will use $\rho = \min\limits_{t > 0} \frac{\vert s(t) \vert}{\vert n(t) \vert}$ instead of ${\rm SNR}$. 

\end{definition}

\section{Main Results} \label{section:main-results}
\subsection{Model of Noise}
\begin{figure}[!htb]
\centering{\includegraphics[scale=0.5]{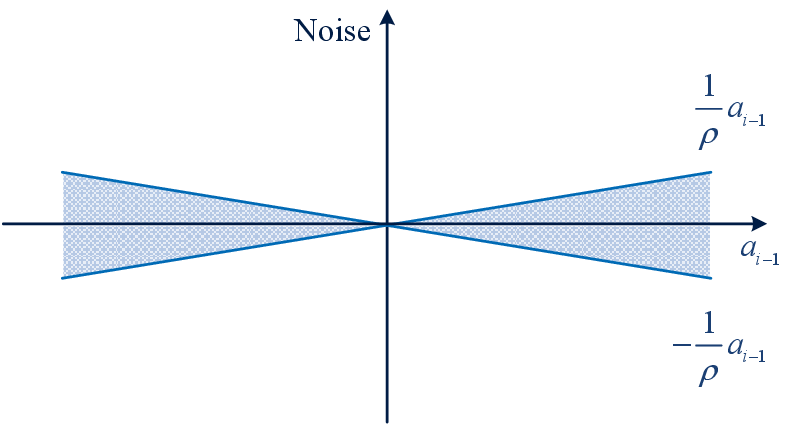}}
\caption{An illustration of the admissible region for the signal noise in the communicated acceleration signal ($\rho = 5$). 
\label{fig_2}}
\end{figure}

\begin{figure}[!htb]
\centering{\includegraphics[scale=0.5]{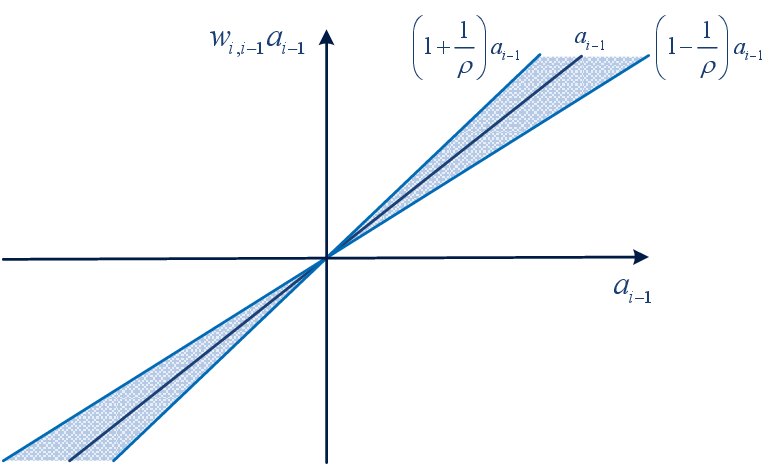}}
\caption{An illustration of the admissible region of the communicated acceleration signal with respect to the actual one ($\rho = 5$).
\label{fig_3}}
\end{figure}

Let $a_{i-1}(t)$ denote the acceleration signal of the preceding vehicle ($(i-1)$-th) and $w_{i, i-1}(t) > 0$ denote the noise factor associated with V2V communication of the acceleration signal of the $(i-1)$-th vehicle to the $i$-th vehicle. Considering the noise affected signal, the communicated acceleration signal available to the $i$-th vehicle is given by $ w_{i,i-1}(t)a_{i-1}(t)$. We define $\rho > 1$ as the signal to noise ratio factor such that $1-\frac{1}{\rho} \leq w_{i, i - 1}(t) \leq 1+\frac{1}{\rho}$, that is, the communicated acceleration signal to the $i$-th vehicle satisfies $\left\vert \left( 1-\frac{1}{\rho} \right) a_{i-1}(t) \right\vert \leq \left\vert w_{i, i-1}(t) a_{i-1}(t) \right\vert \leq \left\vert \left( 1+\frac{1}{\rho} \right) a_{i-1}(t) \right\vert$. An illustration of the admissible region of the noise signal in terms of the ideal communicated acceleration signal is provided in Fig.~\ref{fig_2}, and the range of the communicated signal with respect to the actual signal is provided in Fig.~\ref{fig_3}. 

Considering the communicated acceleration signal is through an $n$-bit communication channel, we model $w_{i,i-1}(t)$ as follows: 
\begin{align}
\label{eq:noise_model}
 w_{i, i - 1}(t) = \left( 1 - \frac{1}{\rho} \right) + \frac{1}{\rho} \left( \sum_{j = 0}^{n - 1} \frac{z_{i, j}(t)}{2^j} \right),   
\end{align}
where $z_{i,j}(t), j = 0,\ldots, n - 1,$ are independent binary random processes. Let $\gamma_{i,j}:= \mathbb{E}[z_{i,j}(t)]$. We assume that $\gamma_{i,j}$ is independent of time, i.e., the characteristics of the noise processes of the communication channel are time-invariant. However, in order to model wide variety of noise processes, we do not assume that $\gamma_{i,j}$'s are known a priori. Note that $\gamma_{i,j} \in (0,1)$ then this noise modeling procedure will allow us to get a communicated signal that takes any value lying between $\left(1-\frac{1}{\rho}\right)a_{i,i-1}$ and $\left(1+\frac{1}{\rho}\right)a_{i,i-1}$.
\subsection{Inter-vehicular Spacing Error Propagation Equation}
Consider the following CTHP control law for the $i$-th vehicle:
\begin{align} \label{eq:control-input-with-noise}
u_{i}(t) = k_a w_{i,i-1}(t) a_{i - 1}(t) - k_v(v_{i}(t) - v_{i - 1}(t)) - k_p \delta_i(t),  
\end{align}
then the governing equation for the $i$-th vehicle is given by:
\begin{align}
    \tau \dddot{x}_i(t) + \ddot{x}_i(t) &= k_a w_{i,i-1}(t) a_{i-1}(t) \nonumber \\
    &\quad - k_v(v_{i}(t) - v_{i - 1}(t)) - k_p \delta_i(t).
\end{align}
The governing equation is stochastic due to the presence of the stochastic noise signal $\omega_{i, i - 1}(t)$; correspondingly, the state is also stochastic. To perform robust string stability analysis, we consider an equivalent deterministic governing equation along the lines described in~\cite{9457141}. By defining the augmented state vector as $\hat{X}(t) = [ x_0(t), v_0(t), a_0(t), x_1(t), v_1(t), a_1(t), \cdots, x_N(t), v_N(t), a_N(t) ]$ and letting $\hat{\omega}(t)$  be the vector of noise signals with its $i$-th component being $\omega_{i, i - 1}(t)$, and the acceleration input to the lead vehicle to be $U(t)$, the augmented state equation for the vehicle platoon can be developed as
\begin{align}
 \dot{\hat{X}}(t) = \hat{A}(\hat{w}(t)) \hat{X}(t) + B U(t),   
\end{align}
where the structures of the matrices $\hat{A}$ and $B$ are given in  ~\cite{9457141}. Through a  discretization process for a sufficiently small time interval, it was shown that 
\begin{align}
 \mathbb{E} [\hat{A}] = \bar{A},  \quad \mathbb{E} [e^{\hat{A}t}] = e^{\bar{A}t},    
\end{align}
where $\bar{A} = \hat{A}(\bar{w})$ with $\bar{w}$ as the expectation of $\hat{w}(t)$. The second equality is true because of the specific lower triangular and banded structure of the governing equations for the CACC case. Let $\bar{X} = \mathbb{E} [\hat{X}]$; in the following, unless specified otherwise, we will use bar to indicate expected value for a corresponding random variable. The evolution of the vehicle states in the platoon converge to the states of the following deterministic state equation:  
\begin{align}
 \dot{\bar{X}}(t) = \bar{A} \bar{X}(t) + B U(t).
\end{align}
Utilizing this procedure, the governing equation for the $i$-th vehicle is given by:
\begin{align}
    \tau \dddot{\bar{x}}_i(t) + \ddot{\bar{x}}_i(t) &= k_a \mathbb{E}[w_{i,i-1}(t)] \bar{a}_{i-1}(t) \nonumber \\
    &\quad - k_v(\bar{v}_{i}(t) - \bar{v}_{i - 1}(t)) - k_p \bar{\delta}_i(t).
\end{align}
Let 
%
\begin{align} \label{eq:katilde1}
 \tilde{k}_a :=  k_a \mathbb{E}[w_{i,i-1}(t)] = \left[ 1 - \frac{1}{\rho} + \frac{1}{\rho} \left( \sum_{j = 0}^{n - 1} \frac{\gamma_{i, j}}{2^j} \right) \right] k_a .   
\end{align}
The governing equation for the $i$-th vehicle can be re-written as:
\begin{align}\label{eq:avg-error-propagation}
    \tau \dddot{\bar{x}}_i(t) + \ddot{\bar{x}}_i(t) &= \tilde{k}_a \ddot{\bar{x}}_{i-1}(t) \nonumber \\
    &\quad - k_v(\dot{\bar{x}}_{i}(t) - \dot{\bar{x}}_{i - 1}(t)) - k_p \bar{\delta}_i(t).
\end{align}
Let $\bar{\delta}_i(t) = \bar{x}_i(t) - \bar{x}_{i-1}(t) + d + h_w \dot{\bar{x}}_i(t)$, and let $\bar{\delta}(s)$ be the Laplace transform of $\bar{\delta}_i(t)$. Then, the inter-vehicular spacing error propagation equation can be derived as 
\begin{align} \label{eq:inter-vehicular-spacing-error-propagation-equation}
 \bar{\delta}_i(s) = \tilde{H}(s; \tau) \bar{\delta}_{i - 1}(s),   
\end{align}
 where $\tilde{H}(s; \tau) = \tilde{\mathcal{N}}(s)/\mathcal{D}(s)$ is the inter-vehicular spacing error propagation transfer function with  
\begin{align*}
\tilde{\mathcal{N}}(s) &= \tilde{k}_a s^2 + k_v s + k_p, \\\mathcal{D}(s) &= \tau s^3 + s^2 + \gamma s + k_p,
\end{align*}
where $\gamma := k_v + h_w k_p$. 

The platoon error propagation equation given by \eqref{eq:avg-error-propagation} was studied in Theorem 1 of  ~\cite{darbha2018benefits} which can be recast for the case considered in this paper as follows:
\begin{theorem} \label{theorem:string-stability-condition-1}
The following are true for the platoon given by the inter-vehicular spacing error propagation equation~\eqref{eq:inter-vehicular-spacing-error-propagation-equation}:
\begin{enumerate}[(a)] 
\item $\| \tilde{H}(j\omega; \tau) \|_{\infty} \le 1$, $\forall \tau \in (0, \tau_0]$, implies that $\tilde{k}_a \in (0,1)$; 
\item given any $\tilde{k}_a \in (0,1)$, $h_w$ satisfying the following: 
\begin{align} \label{eq:hw-lower-bound}
  h_w \ge \frac{2 \tau_0 }{1 + \tilde{k}_a},
 \end{align}
there exist $k_p, k_v > 0$ such that  $\| \tilde{H}(j\omega; \tau) \|_{\infty} \le 1$ for all $\tau \in (0,\tau_0]$.
\end{enumerate}
\end{theorem}

\subsection{Stability Analysis}

\begin{theorem} \label{theorem:string-stability-condition-2}
The following are true for the platoon given by inter-vehicular spacing error propagation equation~\eqref{eq:inter-vehicular-spacing-error-propagation-equation}: 
\begin{enumerate}[(a)] 
\item $\| \tilde{H}(j\omega; \tau) \|_{\infty} \le 1$, $\forall \tau \in (0, \tau_0]$, implies that $\tilde {k}_a \in (0,1)$.
\item given any 
 \begin{align} \label{eq:ka-admissible-range-theorem-3.2}
k_a \in \left(0,\frac{1}{1+\frac{1}{\rho}}\right),
 \end{align} 
and $h_w$ satisfying the following: 
\begin{align} \label{eq:hw-lower-bound1}
  h_w > h_{w,lb}(k_a):= 2 \tau_0 \left( \frac{1 - \left(1-\frac{1}{\rho}\right) k_a}{1 - \left(1+\frac{1}{\rho}\right)^2 k_a^2} \right) 
  ,
 \end{align}
there exist $k_p, k_v > 0$ such that  $\| \tilde{H}(j\omega; \tau) \|_{\infty} \le 1$ for all $\tau \in (0,\tau_0]$;
\item given any $\rho>1$, the minimizing value, $k_a^*$, of $k_a$ and the corresponding minimum value, $h_{w,lb}^*$, of $h_{w,lb}(k_a)$ are given by: 
\begin{align} 
\label{eq:ka-rho}
    k_a^* = \left( \frac{1-\frac{1}{\sqrt{\rho}}}{1+\frac{1}{\sqrt{\rho}}} \right) \frac{1}{\left( 1+\frac{1}{\rho} \right) }, \\
\label{eq:hw-lower-bound2}
  h_{w,lb}^* = \tau_0 \frac{\left(1+\frac{1}{\sqrt{\rho}}\right)^2}{\left( 1+\frac{1}{\rho} \right)}. 
\end{align}
Correspondingly, there exist $k_p, k_v > 0$ such that  $\| \tilde{H}(j\omega; \tau) \|_{\infty} \le 1$ for all $\tau \in (0,\tau_0]$.  
\end{enumerate}
\end{theorem}
\begin{remark}\label{rem:remark1}
    Minimizing $h_{w,lb}$ is useful for improving traffic throughput because a lower time headway can be employed while still guaranteeing ``robust'' string stability. 
\end{remark}

\begin{IEEEproof}
For internal stability, $\gamma - \tau k_p > 0$ by Routh-Hurwitz criterion on $\mathcal{D}(s)$. Applying \emph{Theorem~\ref{theorem:string-stability-condition-1}} to the error propagation equation \eqref{eq:inter-vehicular-spacing-error-propagation-equation}, we infer that:
\begin{enumerate}[(a)]
    \item $\tilde{k}_a \in (0,1)$;
    \item $h_w \ge \dfrac{2\tau_0}{1+\tilde{k}_a}$.
\end{enumerate}

Since $\gamma_{i,j}$'s are not known, we can bound $\tilde{k}_a$ by considering their maximum possible values. Hence, we have 
\begin{align}
\tilde{k}_a \leq \left( 1 - \frac{1}{\rho} + \frac{2}{\rho} \right) k_a = \left( 1+ \frac{1}{\rho} \right) k_a.
\end{align}
Thus, if 
\begin{align} \label{eq:ka-upper-bound}
 k_a < \frac{1}{1 + \frac{1}{\rho}} \implies \tilde{k}_a < 1,   
\end{align}
allowing us to apply part (b) of Theorem \emph{Theorem~\ref{theorem:string-stability-condition-1}}.
%
We then have
\begin{align} \label{eq:hw-lower-bound-without-latency1}
h_{w} \geq \frac{2\tau_0}{1+\tilde{k}_a}.
\end{align}
Again, since $\tilde{k}_a$ is dependent on $\gamma_{i,j}$'s which are not known a priori, we will consider the worst possible lower bound, i.e., the lower bound for $h_w$ that is a maximum over all possible values of $\gamma_{i,j}$'s. To address this, we require  
\begin{align} \label{eq:hw-lower-bound-with-ka1}
 h_w &\ge \max\limits_{\gamma_{i, 0}, \ldots, \gamma_{i, n}} \frac{2 \tau_0}{1 +  \tilde{k}_a}.
\end{align}
Hence, if the above condition holds, then \eqref{eq:hw-lower-bound-without-latency1} will hold. Substituting $\tilde{k}_a$ from \eqref{eq:katilde1} into \eqref{eq:hw-lower-bound-with-ka1}, the condition \eqref{eq:hw-lower-bound-without-latency1} will hold if
\begin{align} \label{eq:hw-lower-bound-with-ka}
 h_w &\ge \max\limits_{\gamma_{i, 0}, \ldots, \gamma_{i, n}} \frac{2 \tau_0}{ 1 + \left( 1 - \frac{1}{\rho} + \frac{1}{\rho} \left( \sum\limits_{j = 0}^{n - 1} \frac{\gamma_{i, j}}{2^j} \right) \right) k_a } \nonumber \\
 &= \frac{2 \tau_0}{\min\limits_{\gamma_{i, 0}, \ldots, \gamma_{i, n}} \left( 1 +  \left( 1 - \frac{1}{\rho} + \frac{1}{\rho} \left( \sum\limits_{j = 0}^{n - 1} \frac{\gamma_{i, j}}{2^j} \right) \right) k_a  \right)  } \nonumber \\
 &= \frac{2 \tau_0}{1 +  \left( 1 - \frac{1}{\rho} \right) k_a}.
\end{align}

In \cite{darbha2018benefits}, the feasible region for the control gains $k_p$ and $k_v$ is specific to one chosen value of $\tilde{k}_a \in (0,1)$. However, in this case, $\tilde{k}_a$ could be any value in the interval $$\mathcal{I}:= \left[\left( 1 - \frac{1}{\rho}\right) k_a, \left( 1 + \frac{1}{\rho}\right) k_a \right]$$ for a given $k_a \in \left(0, \frac{1}{1+\frac{1}{\rho}}\right)$.
Let $\mathcal{F}(\tilde{k}_a)$ denote the set of all $(k_p,k_v)$ that ensure robust string stability for a given value of $\tilde{k}_a$. Unlike in \cite{darbha2018benefits}, we need to show that the set of all feasible $(k_p, k_v)$ that work for any $\tilde{k}_a \in \mathcal{I}$ denoted by $\mathcal{S}$ is non-empty, i.e.,  
$$
\mathcal{S} := \bigcap_{\tilde{k}_a \in \mathcal{I}} \mathcal{F}(\tilde{k}_a) \neq \emptyset.
$$
In the remainder of the proof, we will focus on showing that $\mathcal{S} \neq \emptyset$ and explicitly construct this set for synthesizing the controller. We do so 
 by considering the robust stability condition, $\| \tilde{H}(j \omega; \tau) \|_{\infty}^2 \le 1 \; \forall \tau \in (0, \tau_0]$ and employing a time headway satisfying \eqref{eq:hw-lower-bound-with-ka}. 

According to \emph{Definition~\ref{definition:string-stability}}, robust string stability is guaranteed when $\| \tilde{H}(j \omega; \tau) \|_{\infty}^2 \le 1$, i.e., $\vert \tilde{\mathcal{N}}(j \omega) \vert^2 \le \vert \mathcal{D}(j \omega) \vert^2, \forall \omega$, which can be rewritten as 
 \begin{align}
  \tau^2 \omega^4 + (1 - \tilde{k}_a^2 - 2 \tau \gamma) \omega^2 + \gamma^2 - 2 k_p - k_v^2 + 2 \tilde{k}_a k_p \ge 0.     
 \end{align}
The above condition is satisfied if  $\forall \tau \in (0, \tau_0]$,
 \begin{subnumcases}{}
 1 - \tilde{k}_a^2 - 2 \tau \gamma \ge 0, \\
 \gamma^2 - 2 k_p - k_v^2 + 2 \tilde{k}_a k_p \ge 0,
 \end{subnumcases}
 i.e., 
 \begin{subnumcases}{}
  \gamma \le \frac{1 - \tilde{k}_a^2}{2 \tau_0}, \label{eq:gamma-upper-bound} \\
  \gamma \ge \sqrt{ 2 k_p(1 - \tilde{k}_a) + k_v^2 }. \label{eq:gamma-lower-bound}
 \end{subnumcases}
Thus, by considering the appropriate lower and upper bounds of $\tilde{k}_a$ in the inequalities~\eqref{eq:gamma-upper-bound} and~\eqref{eq:gamma-lower-bound}, these inequalities on $\gamma$ can be satisfied for all $\tilde{k}_a \in \mathcal{I}$ if $\gamma$ satisfies
\begin{subnumcases}{}
 \gamma \le \min\limits_{\tilde{k}_a \in \mathcal{I}} \frac{1 - \tilde{k}_a^2}{2 \tau_0} = \frac{ 1 - \left( 1 + \frac{1}{\rho} \right)^2 k_a^2 }{ 2 \tau_0 }, \label{eq:gamma-upper-bound-min} \\
 \gamma \ge \max\limits_{\tilde{k}_a \in \mathcal{I}} \sqrt{ 2 k_p (1 - \tilde{k}_a) + k_v^2 } \nonumber \\
  \qquad = \sqrt{ 2 k_p \left( 1 - \left( 1 - \frac{1}{\rho} \right) k_a \right) + k_v^2}, \label{eq:gamma-lower-bound-max}
\end{subnumcases}
 from which the admissible range of $\gamma$ is given by 
 \begin{align} \label{eq:gamma-range-CACC-with-noise}
 \sqrt{ 2 k_p \left(1 - \left(1 - \frac{1}{\rho} \right) k_a \right) + k_v^2 } \le \gamma \le \frac{1 - \left(1 + \frac{1}{\rho} \right)^2 k_a^2 }{2 \tau_0}. 
 \end{align}

Next, we will show that the set $\mathcal{S} \ne \emptyset$. First, since $\gamma = k_v + h_w k_p$, 
the right inequality in~\eqref{eq:gamma-range-CACC-with-noise} can be rewritten as 
 \begin{align}
 k_v + h_w k_p \le \frac{ 1 - \left(1 + \frac{1}{\rho} \right)^2 k_a^2 }{2 \tau_0}.
 \end{align}
Thus, an admissible set of $k_p$ and $k_v$ is given by
 \begin{align} \label{eq:kv-kp-feasible-region-1}
 \mathcal{S}_1 := \left\{ (k_p, k_v): k_p > 0, k_v > 0, \frac{k_v}{a_1} + \frac{k_p}{b_1} \le 1 \right\},
 \end{align}
where 
 \begin{align} \label{eq:a1b1}
 \begin{cases}
 a_1 = \dfrac{1 - \left(1 + \frac{1}{\rho} \right)^2 k_a^2 }{2 \tau_0}, \\
 b_1 = \dfrac{ 1 - \left( 1 + \frac{1}{\rho} \right)^2 k_a^2 }{ 2 \tau_0 h_w } = \frac{1}{h_w} a_1.
 \end{cases}
 \end{align}
 Second, 
 the left inequality in~\eqref{eq:gamma-range-CACC-with-noise} can be rewritten as   
 \begin{align}\label{eq:gamma-cond2}
 k_v + h_w k_p \ge \sqrt{ 2 k_p \left( 1 - \left( 1 - \frac{1}{\rho} \right) k_a \right) + k_v^2 }.
 \end{align}
Squaring both sides of \eqref{eq:gamma-cond2} and simplifying, we obtain
\begin{align}
    2 h_w k_v + h_w^2 k_p \ge 2 \left( 1 - \left(1-\frac{1}{\rho}\right) k_a \right).
\end{align}
Thus, an admissible set of $k_p$ and $k_v$ is given by
 \begin{align} \label{eq:kv-kp-feasible-region-2}
 \mathcal{S}_2 := \left\{ (k_p, k_v): k_p > 0, k_v > 0, \frac{k_v}{a_2} + \frac{k_p}{b_2} \ge 1 \right\},
  \end{align} 
 where 
 \begin{align} \label{eq:a2b2}
 \begin{cases}{}
 a_2 = \dfrac{ 1 - \left( 1 - \frac{1}{\rho} \right) k_a  }{ h_w }, \\
 b_2 = \frac{ 2 \left( 1 - \left( 1 - \frac{1}{\rho} \right) k_a \right) }{ h_w^2 } = \frac{2}{h_w} a_2. \end{cases}
 \end{align}

 Then, combining~\eqref{eq:kv-kp-feasible-region-1} and~\eqref{eq:kv-kp-feasible-region-2}, the feasible region of $k_p$ and $k_v$ is given by the set
 \begin{align} \label{eq:kv-kp-region}
 \mathcal{S} = \mathcal{S}_1 \cap \mathcal{S}_2. 
 \end{align}  
Note that $\mathcal{S}$ is nonempty if $a_1 \ge a_2$ or $b_1 \ge b_2$. In particular, notice that  
\[
\frac{a_1}{a_2} = \frac{h_w}{2\tau_0} \left( \frac{1 - \left(1 + \frac{1}{\rho} \right)^2 k_a^2}{1 - \left( 1 - \frac{1}{\rho} \right) k_a  } \right).
\]
Substituting $h_w$ from \eqref{eq:hw-lower-bound1}, we have 
\[
\frac{a_1}{a_2} >1.
\]
 Thus, $\mathcal{S} \neq \emptyset$. This completes the proof of Statement (b) of \emph{Theorem~\ref{theorem:string-stability-condition-2}}. In the following we prove Statement (c) of \emph{Theorem~\ref{theorem:string-stability-condition-2}}.

Let $\bar{h}_w(k_a) := \dfrac{h_{w,lb}(k_a)}{2\tau_0}$, $m:= 1 - \frac{1}{\rho}$, and $n:=\left( 1+\frac{1}{\rho} \right)^2$. Then, 
\begin{align}
  \bar{h}_w(k_a) = \frac{1 - m k_a}{ 1 - n k_a^2 }    
\end{align}
and
\begin{align}
  \frac{d \bar{h}_w(k_a)}{d k_a} = \frac{ - mn k_a^2 + 2 n k_a - m }{(1 - n k_a^2)^2}.    
 \end{align}
Let $f(k_a) := - mn k_a^2 + 2 n k_a - m$. Then, the roots of $f(k_a) = 0$ are given by 
%
%
%
\[
r_{1,2} := \frac{n \mp \sqrt{ n (n - m^2) }}{m n}.
\]
Note that $r_1 < \frac{1}{1+\frac{1}{\rho}}$ and $r_2 > \frac{1}{1+\frac{1}{\rho}}$. The function $f(k_a)$ vs. $k_a$ is provided as an illustration for $\rho = 5$. 
\begin{figure}[!b]
\centering{\includegraphics[scale=0.6]{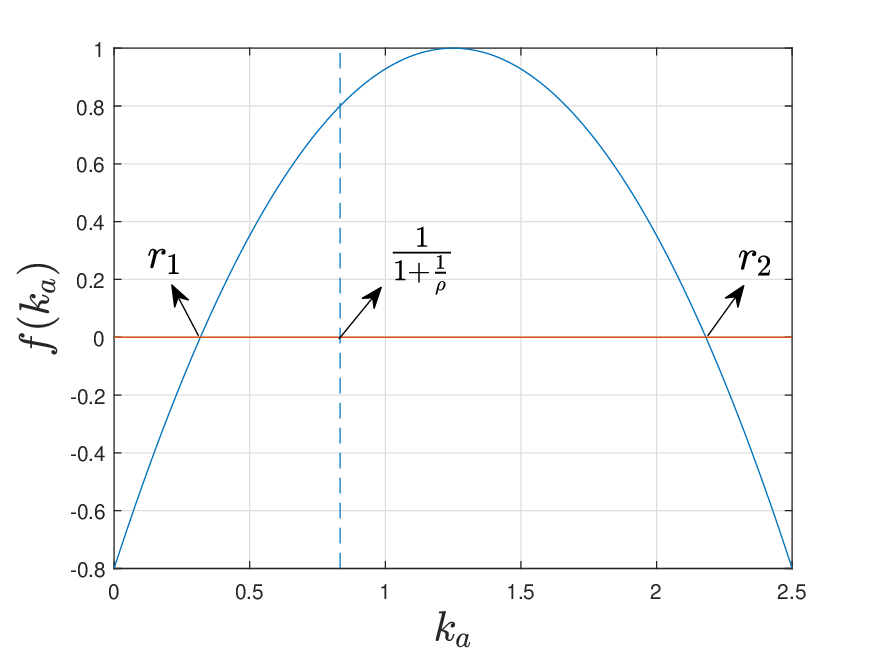}}
\caption{$f(k_a)$ vs. $k_a$ when $\rho = 5$.
\label{fig:optimal-ka}}
\end{figure}
Since $k_a \in \left(0, \frac{1}{1 + \frac{1}{\rho}}\right)$, we need to consider only $r_1$. Hence, we have 
%
\begin{align}
 \begin{cases}
  \dfrac{d \bar{h}_w}{d k_a} < 0, \mbox{when}~k_a \in (0, r_1); \\
   \dfrac{d \bar{h}_w}{d k_a} > 0, \mbox{when}~k_a \in \left(r_1, \frac{1}{1 + \frac{1}{\rho}}\right).
 \end{cases}   
\end{align}
Thus, 
 \begin{align}
  \min\limits_{k_a \in \left(0, \frac{1}{1 + \frac{1}{\rho}}\right)} \bar{h}_w(k_a) = \bar{h}_w(r_1).    
 \end{align}
Substituting $m$ and $n$ into $r_1$, we obtain \
\begin{align}
k_{a}^{*} := r_1 
 &= \left( \frac{ 1 - \frac{1}{\sqrt{\rho}} }{ 1 + \frac{1}{\sqrt{\rho}} } \right) \left( \frac{1}{1 + \frac{1}{\rho}} \right).
\end{align}
Further, substituting $r_1$ into $\bar{h}_w(r_1)$, we obtain 
 \begin{align}
  \bar{h}_{w}(r_1) = \frac{\left( 1 + \frac{1}{\sqrt{\rho}} \right)^2}{2 \left(1 + \frac{1}{\rho}\right)}.   
 \end{align}
Therefore, 
 \begin{align}
  h_{w,lb}^{*} = 2 \tau_0 \bar{h}_w(r_1) = \tau_0 \frac{ \left(1 + \frac{1}{\sqrt{\rho}}\right)^2 }{1 + \frac{1}{\rho}}.   
 \end{align}
This completes the proof of Statement (c) of \emph{Theorem~\ref{theorem:string-stability-condition-2}}.


Finally, we will prove the lower bound of the time headway given by~\eqref{eq:hw-lower-bound2} satisfies the internal stability condition. 
For this purpose, note that 
\begin{align} 
  h_w k_p \ge \tau_0 k_p \frac{\left(1+\frac{1}{\sqrt{\rho}}\right)^2}{\left( 1+\frac{1}{\rho} \right)}. 
 \end{align}
Thus, $\gamma = k_v + h_w k_p > h_w k_p \ge \tau_0 k_p$.

%
%
Therefore, this completes the proof of \emph{Theorem~\ref{theorem:string-stability-condition-2}}.    
\end{IEEEproof}





\begin{remark}
    If $\rho \to \infty$ (no signal noise case), then $k_a \in (0,1)$ and $h_w = \frac{2 \tau_0}{1 + k_a}$ as given in in~\cite{darbha2018benefits}. 
\end{remark}

 \begin{remark}
In~\eqref{eq:hw-lower-bound1}, if $\rho \to \infty$, then the lower bound of the time headway reduces to $h_w \ge h_{w, lb}(k_a) = \frac{2 \tau_0}{1 + k_a}$ as given  in~\cite{darbha2018benefits}. For the same $k_a$ value, the lower bound of the time headway increases as $\rho$ decreases 
in the presence of signal noise. In addition, according to~\eqref{eq:ka-admissible-range-theorem-3.2}, the upper bound of $k_a$ is smaller when compared to the no noise case, which also factors into the increase of the lower bound of the time headway in the presence of noise. In addition, in~\eqref{eq:hw-lower-bound2}, if $\rho \to \infty$, the minimum lower bound of the time headway reduces to $h^{\ast}_{w,lb} \to \tau_0$, and the corresponding optimal $k_a$ given by~\eqref{eq:ka-rho} becomes $k_a^{\ast} \to 1$.  
 \end{remark}

\section{Numerical Simulations} \label{section:numerical-simulations} 
In this section, we present a numerical example and simulation results to corroborate the results in Section~\ref{section:main-results}. We consider the following numerical values for the system parameters: $\tau_0 = 0.5$ s, $d = 5$ m, $N = 12$. In the numerical simulation, $\tau$ was chosen as $\tau = \tau_0$. We assume that the lead  vehicle experiences an external disturbance which causes a perturbation on its acceleration, denoted as $a_0(t)$, as follows: 
\begin{align} \label{eq:a0-expression-simulation}
 a_0(t) = 
 \begin{cases}
  0.5 \sin (0.1 ( t - 10 )), 10 ~ {\rm s} < t < (10 + 20 \pi) ~ { \rm s}, \\
  0, \mbox{otherwise},
 \end{cases}   
\end{align}
and under which the performance of the communication and control strategy considered in Section~\ref{section:main-results} will be evaluated in the following.





\begin{figure}[!htb]
\centering{\includegraphics[width=\hsize]{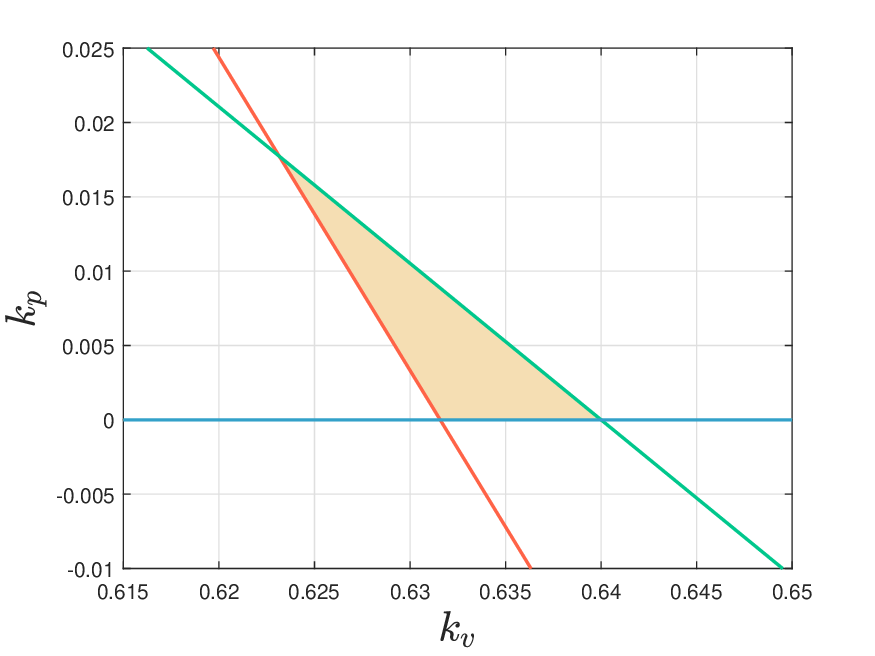}}
\caption{The feasible region of $k_p$ and $k_v$ when $k_a = 0.5$, $h_w = 0.95$ s.
\label{fig:kv-kp-region}}
\end{figure}

\begin{figure}[!htb]
\centering{\includegraphics[width=\hsize]{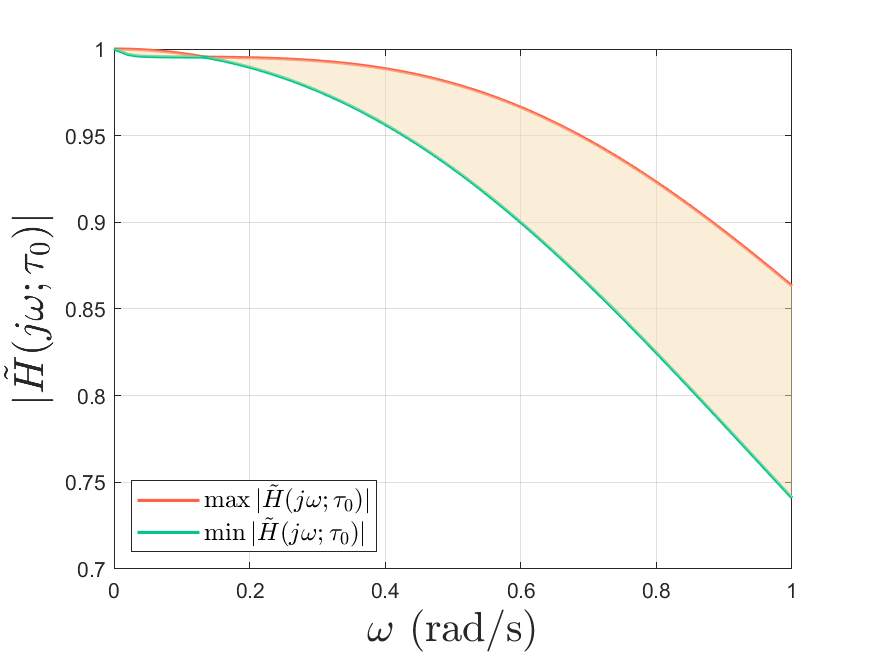}}
\caption{$\vert \tilde{H}(j \omega; \tau_0) \vert$ when $k_a = 0.5$, $k_p = 0.009$, $k_v = 0.63$, $h_w = 0.95$ s.
\label{fig:H-tilde-norm-hw-0point75}}
\end{figure}

Assume $\rho = 5$, then the upper bound of $k_a$ can be computed as $k_a < \frac{1}{1 + \frac{1}{\rho}} = 0.8333$. Then, by choosing $k_a = 0.5$ and substituting $\tau_0, \rho, k_a$ values in~\eqref{eq:hw-lower-bound1}, $h_w \ge 0.9375$ s. Choosing $h_w = 0.95$ s, the feasible region of $k_p$ and $k_v$ is obtained as shown in Fig.~\ref{fig:kv-kp-region}, from which we choose $k_p = 0.009$, $k_v = 0.63$ for the numerical simulations. 
With the above chosen values of the time headway and control gains, the frequency response of $\left\vert \tilde{H}(j \omega; \tau_0) \right\vert$ is shown in Fig.~\ref{fig:H-tilde-norm-hw-0point75} which demonstrates string stability of the platoon. 
 
\begin{figure}[!htb]
\centering{\includegraphics[width=\hsize]{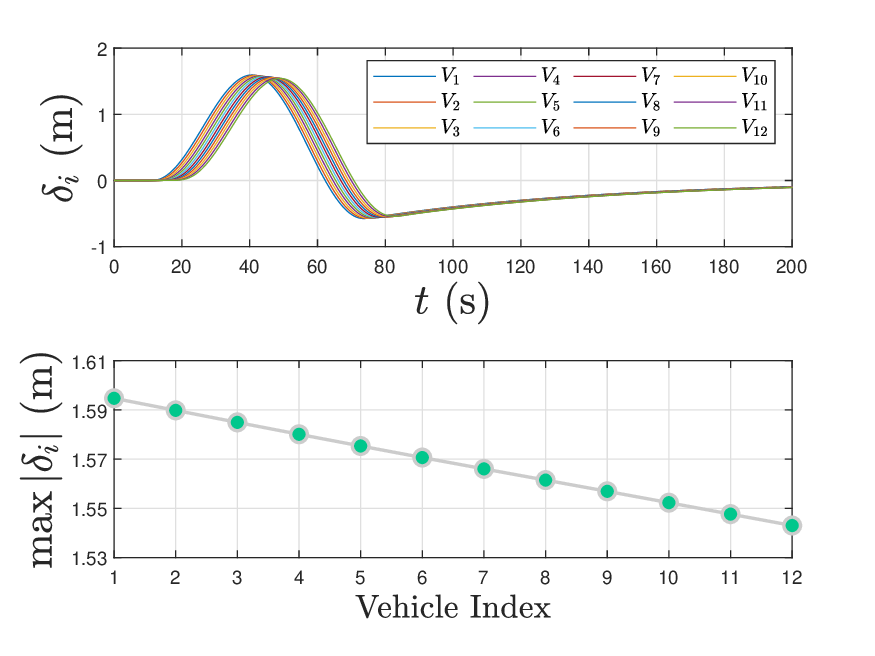}}
\caption{The inter-vehicular spacing errors of the following vehicles ($k_a = 0.5$, $k_p = 0.009$, $k_v = 0.63$ and $h_w = 0.95$ s).
\label{fig:delta-hw-0point75}}
\end{figure}

\begin{figure}[!htb]
\centering{\includegraphics[width=\hsize]{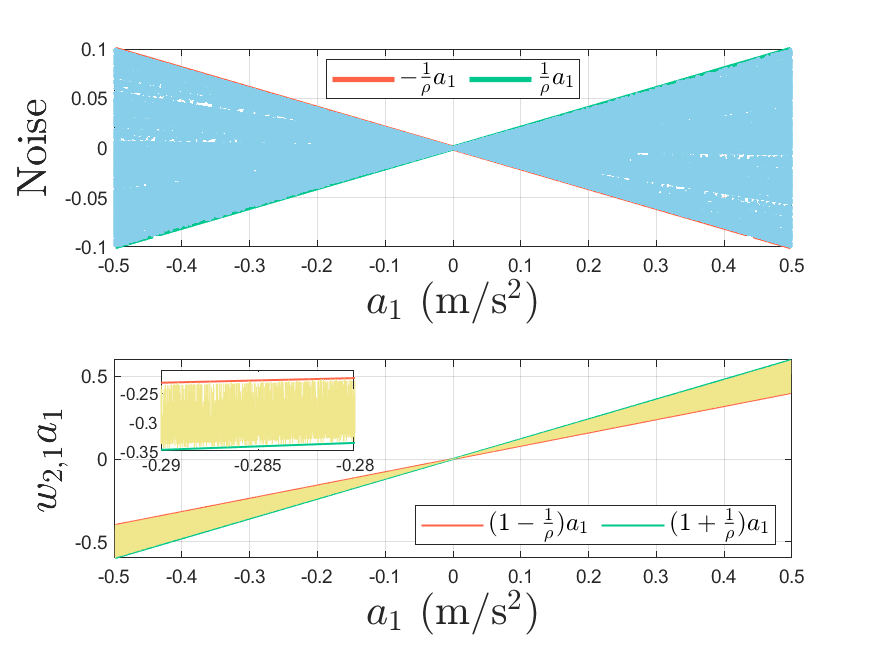}}
\caption{The noise and the communicated acceleration signal from vehicle 1 to vehicle 2 ($k_a = 0.5$, $k_p = 0.009$, $k_v = 0.63$ and $h_w = 0.95$ s).
\label{fig:noise-and-communicated-signal-hw-0point95}}
\end{figure}

Suppose we model noise in \eqref{eq:noise_model} by choosing $n = 16$ discretizations, and the expectation of the binary random variables $z_{i, j}$ are respectively: $\gamma_{i, 0} = 0.8055$, $\gamma_{i, 1} = 0.5767$, $\gamma_{i, 2} = 0.1829$, $\gamma_{i, 3} = 0.2399$, $\gamma_{i, 4} = 0.8865$, $\gamma_{i, 5} = 0.0287$, $\gamma_{i, 6} = 0.4899$, $\gamma_{i, 7} = 0.1679$, $\gamma_{i, 8} = 0.9787$, $\gamma_{i, 9} = 0.7127$, $\gamma_{i, 10} = 0.5005$, $\gamma_{i, 11} = 0.4711$, $\gamma_{i, 12} = 0.0596$, $\gamma_{i, 13} = 0.6820$, $\gamma_{i, 14} = 0.0424$, $\gamma_{i, 15} = 0.0714$. 
The evolutions of the inter-vehicular spacing errors are shown in Fig.~\ref{fig:delta-hw-0point75}.  
In addition, as an example, the noise and the communicated acceleration signal from vehicle 1 to vehicle 2 are shown in Fig.~\ref{fig:noise-and-communicated-signal-hw-0point95}.

\begin{figure}[!htb]
\centering{\includegraphics[width=\hsize]{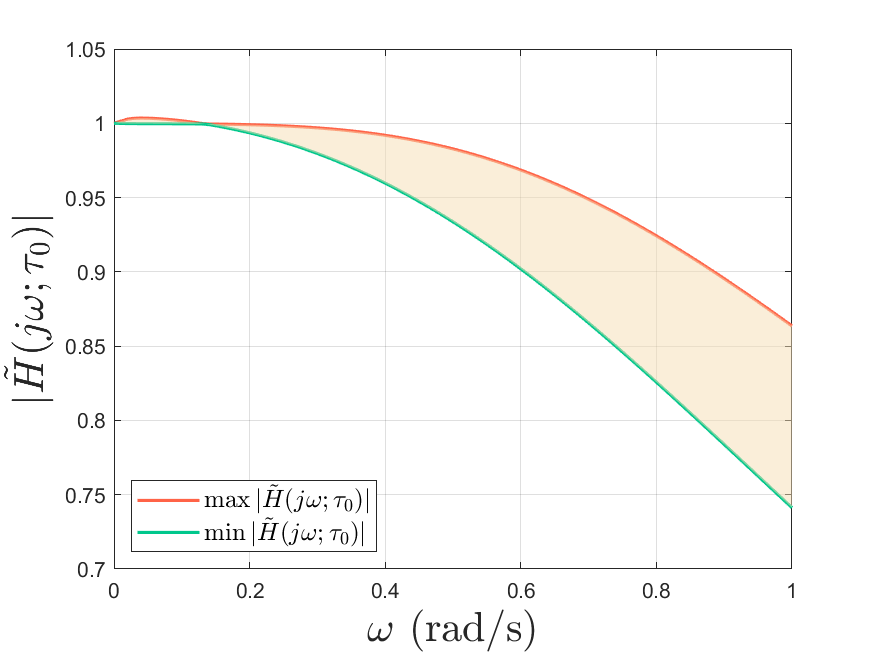}}
\caption{$\vert \tilde{H}(j \omega; \tau_0) \vert$ when $k_a = 0.5$, $k_p = 0.009$, $k_v = 0.63$, $h_w = 0.65$ s.
\label{fig:H-tilde-norm-hw-0point65}}
\end{figure}

\begin{figure}[!htb]
\centering{\includegraphics[width=\hsize]{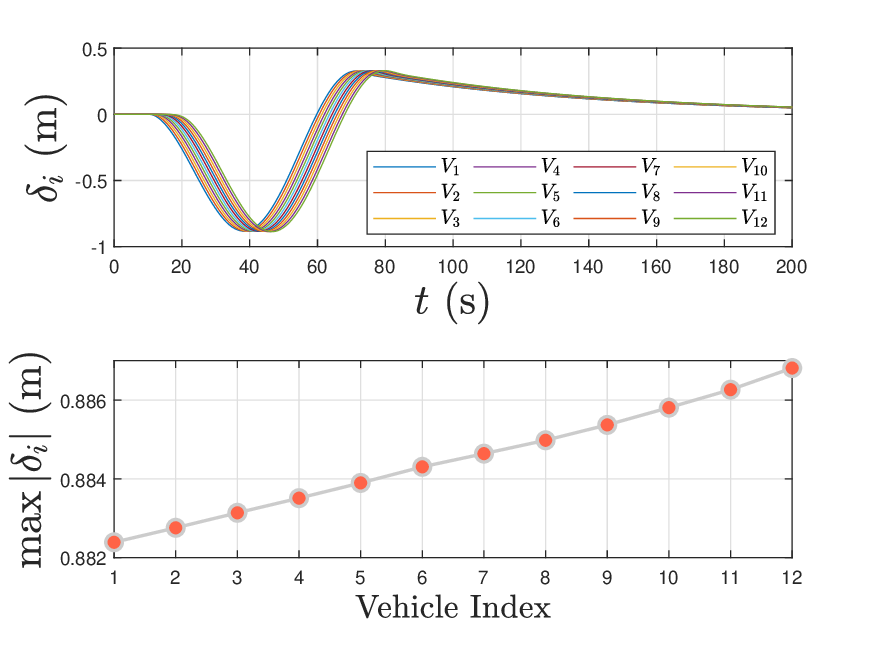}}
\caption{The inter-vehicular spacing errors of the following vehicles ($k_a = 0.5$, $k
_p = 0.009$, $k_v = 0.63$ and $h_w = 0.65$ s).
\label{fig:delta-hw-0point65}}
\end{figure}

For comparison, under the same condition as above except for $h_w = 0.65$ s, 
then the frequency response of $\vert \tilde{H}(j \omega; \tau_0) \vert$ is shown in Fig.~\ref{fig:H-tilde-norm-hw-0point65} which exhibits string instability. In addition, the inter-vehicular spacing errors 
of the following vehicles are shown in Fig.~\ref{fig:delta-hw-0point65}. 

\begin{figure}[!htb]
\centering{\includegraphics[width=\hsize]{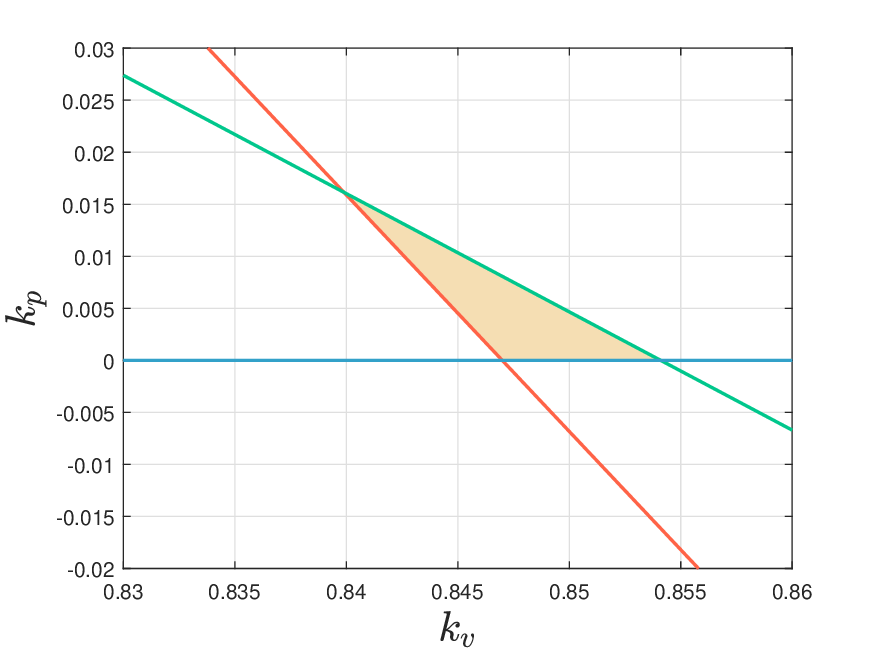}}
\caption{The feasible region of $k_p$ and $k_v$ when $k_a = k_a^{\ast}$, $h_w = 0.88$ s.
\label{fig:kv-kp-region-2}}
\end{figure}

\begin{figure}[!htb]
\centering{\includegraphics[width=\hsize]{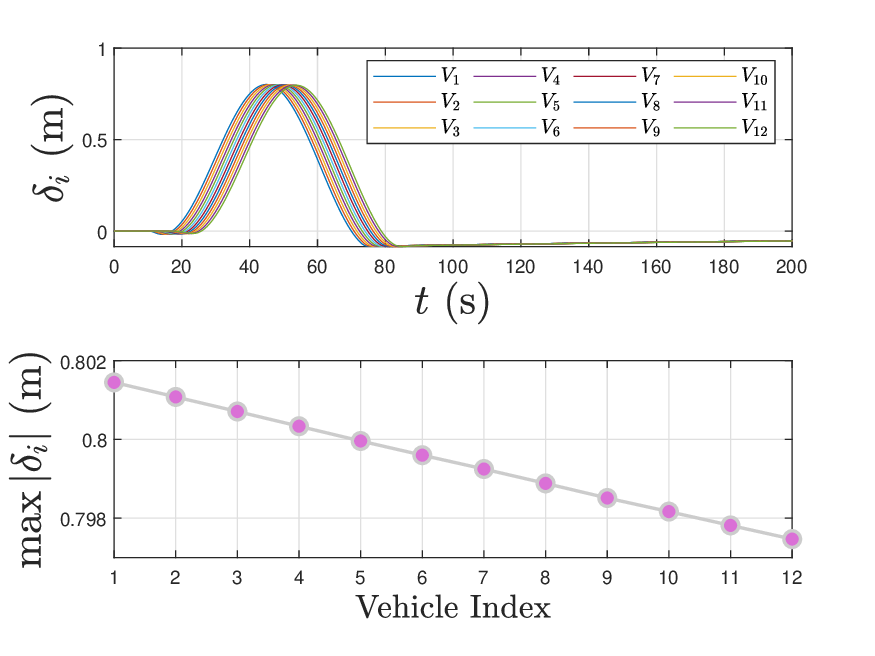}}
\caption{The inter-vehicular spacing errors of the following vehicles ($k_a = k_a^{\ast}$, $k
_p = 0.003$, $k_v = 0.85$ and $h_w = 0.88$ s).
\label{fig:delta-optimal}}
\end{figure}

\begin{figure}[!htb]
\centering{\includegraphics[width=\hsize]{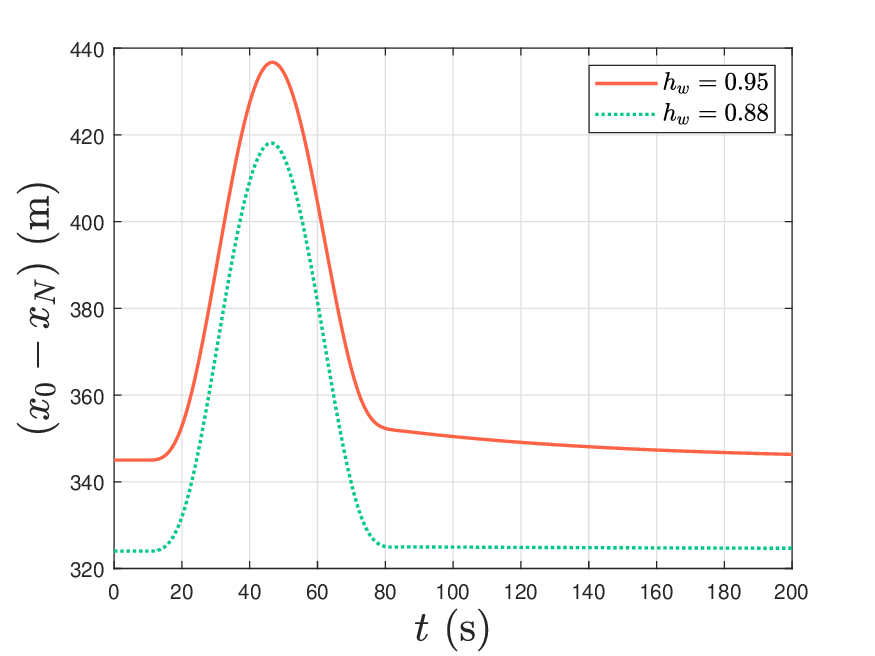}}
\caption{The comparison of the length of the platoon between $h_w = 0.95$ and $h_w = 0.88 \approx h^{\ast}_{w, lb}$.
\label{fig:comparison-of-length}}
\end{figure}

In addition, we have conducted numerical simulations to evaluate the performance of the platoon for $k_a^{\ast}$ and $h_{w,lb}^\ast$ given in Statement (c) of Theorem~\ref{theorem:string-stability-condition-2}. For $\rho=5$, we can obtain $k_a^{\ast} = 0.3183$ and $h^{\ast}_{w,lb} = 0.8727$. Choosing $k_a = k_a^{\ast}$ and $h_w = 0.88$, the feasible region of $k_v$ and $k_p$ is shown in Fig.~\ref{fig:kv-kp-region-2}.  
Choosing $k_p = 0.003$, $k_v = 0.85$ from the feasible region, the evolution of the inter-vehicular spacing errors is provided in Fig.~\ref{fig:delta-optimal}. 
Figure~\ref{fig:comparison-of-length} provides the comparison of the platoon size ($x_0 - x_N$) for the two time headways, $h_w = 0.95$ s and $h_w = 0.88$ s, indicating higher throughput as stated in Remark~\ref{rem:remark1}. 

From the numerical simulation results, it can be seen that this vehicle platoon is robustly string stable with the synthesized control gains and time headway according to the analysis and design procedure provided in  Section~\ref{section:main-results}.  


\section{Conclusion} \label{section:conclusion}
We have investigated robust string stability of connected and autonomous vehicle platoons with cooperative adaptive cruise control systems subject to noisy V2V communication.  
 We have derived conditions on control gains for predecessor acceleration, relative velocity and spacing errors and a lower bound for time headway that are dependent on the signal to noise ratio due to V2V communication of acceleration, while ensuring that the CACC system is internal and string stable.   
 We have provided a systematic analysis through which one can select control gains and time headway for a given SNR. 
 We have also provided an illustrative example and corresponding simulation results to demonstrate the main results. 
For future work, we will investigate 
extension to noisy communication when information from multiple predecessors is employed. 
\appendices

\bibliographystyle{IEEEtran}

\bibliography{ref}

\end{document}